# Five-second coherence of a single spin with single-shot readout in silicon carbide

Christopher P. Anderson[1,2,†], Elena O. Glen[1,†], Cyrus Zeledon[1], Alexandre Bourassa[1], Yu Jin[3], Yizhi Zhu[1], Christian Vorwerk[1], Alexander L. Crook[1,2], Hiroshi Abe[4], Jawad Ul-Hassan[5], Takeshi Ohshima[4], Nguyen T. Son[5], Giulia Galli[1,3,6], David D. Awschalom[1,2,6,*]

[1]Pritzker School of Molecular Engineering, University of Chicago, Chicago, Illinois 60637, USA.
[2]Department of Physics, University of Chicago, Chicago, Illinois 60637, USA.
[3]Department of Chemistry, University of Chicago, Chicago, Illinois 60637, USA.
[4]National Institutes for Quantum Science and Technology, 1233 Watanuki, Takasaki, Gunma 370-1292, Japan.
[5]Department of Physics, Chemistry and Biology, Linköping University, SE-581 83 Linköping, Sweden.
[6]Center for Molecular Engineering and Materials Science Division, Argonne National Laboratory, Lemont, IL 60439, USA.
† These authors contributed equally to this work.
[*]email:awsch@uchicago.edu

**Abstract**
An outstanding hurdle for defect spin qubits in silicon carbide (SiC) is single-shot readout—a deterministic measurement of the quantum state. Here, we demonstrate single-shot readout of single defects in SiC via spin-to-charge conversion, whereby the defect's spin state is mapped onto a long-lived charge state. With this technique, we achieve over 80% readout fidelity without pre- or post-selection, resulting in a high signal-to-noise ratio (SNR) that enables us to measure long spin coherence times. Combined with pulsed dynamical decoupling sequences in an isotopically purified host material, we report single-spin $T_2>5$ s, over two orders of magnitude greater than previously reported in this system. The mapping of these coherent spin states onto single charges unlocks both single-shot readout for scalable quantum nodes and opportunities for electrical readout via integration with semiconductor devices.

**Introduction**

Solid-state defect spins hold promise for use in quantum information processing, sensing, and communication because of their unique combination of long coherence times (*1–4*), a spin-photon interface (*5, 6*), and the availability of nuclear registers for use as robust quantum memories (*7, 8*). The neutral divacancy ($VV^0$) in SiC boasts these features with the added advantages of the SiC material platform, including wafer-scale commercial availability, CMOS compatibility, and the ability to fabricate hybrid photonic (*9, 10*), electrical (*11*), and mechanical devices (*12, 13*).

Typically, for optically-active defects spins, single-shot readout is performed through spin-dependent fluorescence probed with narrow-line lasers resonant with the defect's optical cycling transitions (*14*). However, this method suffers from spin-flip errors due to non-unity branching ratios in the defect's optical excited state (*5*). As a result, only a finite number of spin-correlated photons are scattered before destroying the state. Combined with poor collection efficiencies, the number of measured photons per shot is usually low ($N \ll 1$) unless photonic devices are used to enhance emission and collection. As such, a key hurdle for the divacancy system to date has been the ability to perform single-shot readout of the defect's spin state (*15*). This single-shot readout unlocks the ability to perform the entanglement distribution (*16*) and quantum error correction (*17*) needed to make quantum networks a reality, and provides an increased signal-to-noise ratio for quantum sensing.

Another avenue towards single-shot readout is spin-to-charge conversion (SCC), which maps the defect spin state onto a robust, long-lived charge state. For isolated single defect spins, SCC is an all-optical technique that has been used to achieve high-fidelity single-shot readout, but has thus far been limited to demonstrations using the $NV^-$ center in diamond (*18, 19*). In this work, we demonstrate the first ever implementation of SCC for $VV^0$ in SiC by performing spin-selective ionization followed by all-optical single-shot readout of the charge state. Using this technique, we can determine an initially prepared spin state with over 80% fidelity. We note that in this work, we do not pre- or post-select the charge state or resonance condition, a common strategy used to artificially boost fidelity at the cost of success rate (*14, 18, 19*). Critically, we also achieve this readout in the absence of photonic enhancement, demonstrating the value of SCC to other systems where nanofabrication remains a challenge, or where detector technologies may be limited. Additionally, we deepen our understanding of the SCC process for the $VV^0$ defect with *ab initio* DFT calculations that clarify the charge-transition process, which combined with theoretical modelling, assist with further optimization of this type of readout.

The high-fidelity and single-shot readout provided by the SCC technique enable us to perform experiments that otherwise would be infeasible due to low SNR, such as when the time per experimental 'shot' becomes prohibitively long. With this advantage, we employ the SCC technique to probe the unexplored limits of the spin lifetime and coherence time for $VV^0$. These times fundamentally determine the divacancy's performance in future quantum architectures by limiting metrics such as quantum memory times and sensitivity for ac sensing schemes (*20–22*). Here, we first establish an experimentally-limited lower bound on the spin $T_1$ time of at least 103 seconds for $VV^0$, over two orders of magnitude longer than previously reported (*7*). In this isotopically purified sample, we combine the reduction of the defect's noisy nuclear environment with the use of dynamical decoupling sequences to preserve coherence (*23, 24*). As a result, we measure a $T_2$ time of $5.3 \pm 1.3$ seconds, over three orders of magnitude greater than the natural

Hahn-echo $T_2$ time (*3*). These metrics establish VV$^0$ as a premier system with coherence times that exceed previous reports for electron spin qubits in both natural and highly isotopically purified silicon (*25, 26*), diamond (*27, 28*), and SiC (*2, 29*).

The results presented in this work develop SiC-based systems as a promising platform for quantum technologies, where both deterministic readout of the spin state and long coherence times are necessary for heralded entanglement generation, high gate fidelities, and the development of network components such as quantum repeaters. This work also opens avenues that utilize the CMOS compatibility of SiC for the integration of electron spin-based systems in classical electrical devices that are sensitive to single charges.

**Results**
### Optical and charge transitions of the divacancy in SiC
The neutral divacancy in 4H-SiC is a deep-level defect consisting of a carbon and silicon vacancy pair. The dangling bonds from atoms neighboring these vacancies form a spin-triplet ground state (*30*) with spin sublevels that can be polarized and read out with laser light and manipulated with microwaves (Fig. 1A). In this work, we use laser light resonant with the E$_x$ and E$_{1,2}$ spin-selective optical transitions ('resonant excitation'), corresponding to the $m_s=0$ and $m_s=\pm 1$ spin sublevels (*5*), respectively (Fig. 1B). The divacancy spin state can be efficiently initialized to $m_s=0$ via excitation of the E$_{1,2}$ transition, which depletes the $m_s=\pm 1$ population through optical pumping and also non-radiatively polarizes into $m_s=0$ via a spin-singlet intersystem crossing (*31, 32*). In past work, spin-photon readout is performed by collecting photoluminescence (PL) scattered when pumping on one of the more cycling resonant optical lines, such as the E$_x$ transition. However, when pumping on a single optical transition, spin-flips from the excited state (*31*) cause a depopulation that prevents indefinite optical readout. The finite number of photons emitted before destroying the state ultimately limits the fidelity of the spin-photon readout technique.

As an alternative to readout via the spin-photon interface, the divacancy hosts robust charge states (*33*) that can also be manipulated and read out using laser light (*11, 34, 35*). In this work, we use the fact that non-neutral states of the divacancy do not appreciably photoluminesce under resonant excitation that is tuned to the neutral state's zero-phonon line (ZPL). When the optical lines corresponding to both $m_s=0$ and $m_s=\pm 1$ of VV$^0$ are simultaneously pumped ('charge readout'), the emitted PL does not reflect the spin state but rather if the defect is in the neutral state or not, provided the lasers are on resonance. Thus, a reduction in PL distinguishes 'dark' ionized states from the 'bright' neutral state (Fig. 1C), as the optical lines are stable in this sample. For the VV$^0$ in SiC, this dark state has been established as the negatively charged divacancy (VV$^-$) (*11, 34–36*). Crucially, probing the divacancy charge state with this light is not energetic enough to convert VV$^-$ to VV$^0$ and vice versa via a direct one-photon process (Fig. 1A). The result is that non-destructive measurement of the charge state of the defect is possible with high fidelity. In this work, we also rely on deterministic preparation of the defect into the neutral charge state. Previous reports have shown that laser light above 1.3 eV resets the charge state from VV$^-$ to VV$^0$, and that light around 705 nm (1.76 eV) is extremely efficient in charge initializing the divacancy to its neutral state (VV$^0$) (*11*). However, the fidelity of this process has remained unexplored to date.

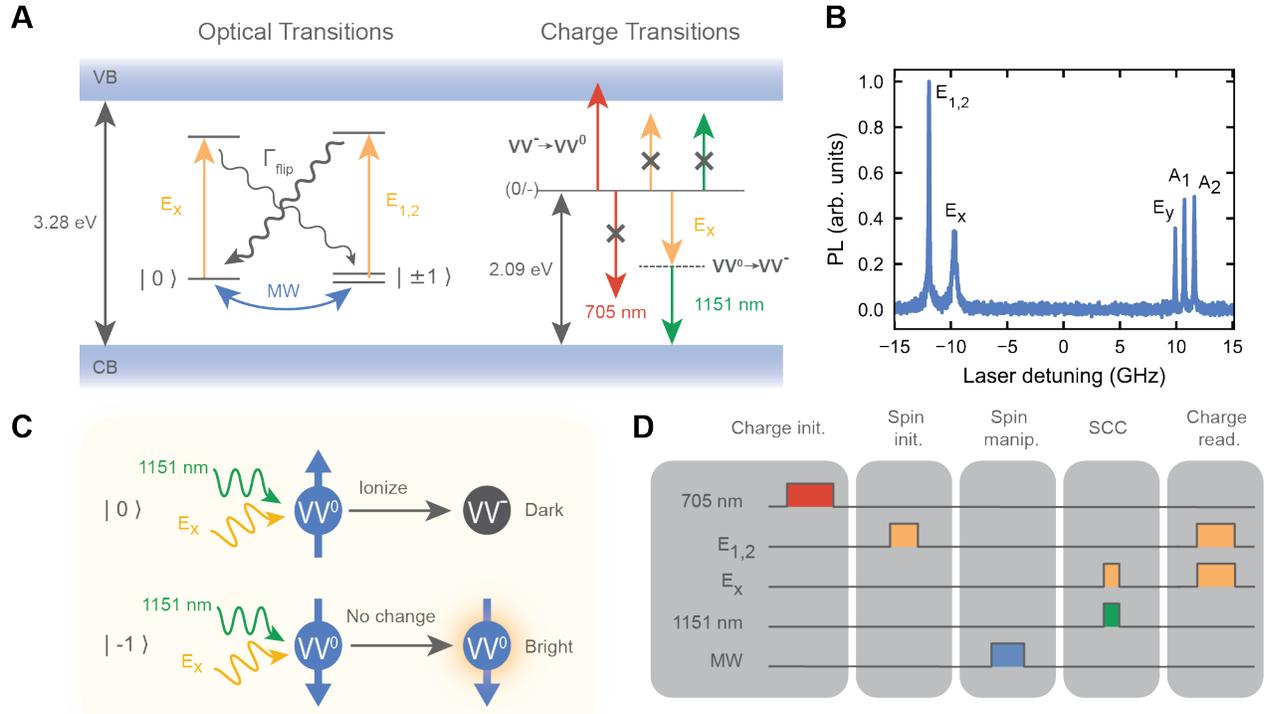

**Fig. 1. Control and readout of spin and charge states of the divacancy.** (**A**) Optical and charge transitions of the divacancy. Excitation of the $E_x$ and $E_{1,2}$ spin-selective optical transitions is performed using ~1131 nm light (yellow). Spin flips ($\Gamma_{flip}$) prevent indefinite readout of the spin state with laser pumping. Microwave manipulation (MW) is used to induce ground state spin-sublevel transitions. SCC is performed by excitation of the $E_x$ transition followed by an ejection of a hole by the 1151 nm ionization laser (green). The dashed line represents the $VV^0$ excited state. The 705 nm light (red) resets the divacancy from $VV^-$ to $VV^0$. The individual lasers used for SCC and charge resetting do not have enough energy to induce other charge transitions via a one-photon process (denoted by a 'X'). (**B**) Photoluminescence excitation (PLE) spectrum of a single divacancy reveals its six spin-selective optical transitions at T=5 K, B=18 G with continuous microwave driving of the $m_s=0 \leftrightarrow m_s=-1$ transition. Detuning is relative to a center laser frequency of 265.1408 THz and the transverse strain splitting is 9.78 GHz. (**C**) Mapping of the spin state onto the charge state. Pumping of the $E_x$ transition allows for ionization of $m_s=0$ with the 1151 nm laser. The SCC step is followed by charge readout via pumping of both the $E_x$ and $E_{1,2}$ transitions. Detection of PL signifies that the divacancy is in its bright, neutral (dark, ionized) state, and therefore was prepared to $m_s=-1$ ($m_s=0$). (**D**) Typical experimental pulse sequence. After the charge and spin initialization, and MW manipulation of the spin state, single-shot readout of the spin state is performed with SCC followed by readout of the charge state.

Mapping of the spin state onto the charge state (SCC) via a spin-selective two-photon ionization process provides us with an avenue towards performing high-fidelity, deterministic measurement of the spin state via readout of the charge state. Here, we access the defect's excited state in a one-photon, spin-selective manner using a narrow-line laser tuned to one of the resolved optical transitions (Fig. 1A, 1B). A second 'ionization laser' (1151 nm) takes the defect from its excited state to the ionized state ($VV^-$) via a one-photon process by ejecting a hole (*37*). We select the ionization laser wavelength based on the results of DFT calculations we perform for the $VV^0$ charge transition energies, which for the (0/-) transition is calculated to be 2.09 eV, in good agreement with previous work (*30, 37*). The ionization laser (1151 nm) is red-detuned from the defect's ZPL (1131 nm) so as not to excite any optical transitions, while still providing enough energy to ionize the defect from the excited state, as the combined energy of these photons (2.17 eV) enables the 2.09 eV (0/-) charge transition (*33, 37*) to occur (Fig. 1A). Specifically, we go beyond this estimate based on the charge transition levels and directly compute using DFT the energy required to go from the VV$^0$ optical excited state to the VV$^-$ state with a hole at the valence-band maximum. Our DFT calculations show that photoionization from the excited state requires ~1.03 eV, an energy that is slightly less than the calculated ZPL of 1.196 eV, in good agreement with experimental results (Supplementary Materials). This means that we can use a narrow, low-power resonant laser tuned to a single optical transition to provide spin-selectivity alongside a high-power, red-detuned ionization tone to induce, in total, a two-photon, spin-dependent ionization of the defect.

Once we have spin-selectively mapped the spin onto the defect's charge state with the SCC step, we can then perform single-shot readout of the charge state by addressing the defect with both $E_x$ and $E_{1,2}$ resonant light and collecting PL (Fig. 1C). Thus, we are equipped with a full suite of techniques to initialize the charge and spin state on-demand, manipulate the spin with microwave pulses, convert the spin state to a charge state, and perform charge readout with these lasers and controls (Fig. 1D).

### Charge control and readout

We first demonstrate the robustness of the divacancy charge state and our ability to perform single-shot, high-fidelity optical readout of this state. We characterize the longevity of the VV$^0$ charge state with a sequence consisting of a 705 nm charge initialization pulse, a variable delay, and an optical charge readout from which we extract a charge lifetime $\tau_{ch}$ =6.9±0.9 s (Fig. 2A) (Supplementary Materials). The finite duration of the charge lifetime when the defect is not under illumination is likely due to diffusion of charges from nearby shallow nitrogen donors, as the material is slightly n-type (Materials and Methods). The charge lifetime is a critical timescale that dictates the longest permissible time during an experiment where the neutral defect, and therefore the spin, remains stable. Thus, the charge lifetime ultimately limits the qubit's lifetime and is a stringent cutoff for sensing and memory applications. Fortunately, previous work has shown much longer charge lifetimes (*34*), where future tuning of the Fermi level or balance of deep traps in the material may extend our measured timescale by many orders of magnitude. Interestingly, this charge instability is also linked to our ability to optically ionize the defect on-demand, representing a tradeoff to be optimized in future materials design. Our measured charge lifetime, however, is still many orders of magnitude longer than the saturated spin-flip lifetime, which we measure to be 3.3±0.1 $\mu s$ (Supplementary Materials), resulting in a longer possible readout window and more scattered photons before destroying the state.

We next demonstrate our ability to perform single-shot readout of the charge state. First, we either charge-initialize to the bright state using a long, 705 nm laser pulse or spin-agnostically initialize to the dark state using an 'ionization pulse' where both $E_x$ and $E_{1,2}$ resonant lasers and the 1151 nm ionization laser are simultaneously turned on. This preparation into either the bright or dark charge state is followed by optical charge readout. Fig. 2B displays the number distribution of photons collected during the charge readout step for both the prepared bright and dark initial state. We calculate that for preparation into the bright (dark) state the mean photon number is $N$=100±1 ($N$=1.3±1.1) (Supplementary Materials). We determine that for a cutoff of $N$=4 photons, the fidelity, $F_{charge}$ (*38*) is maximized at 98.7±1.3% (Supplementary Materials), representing the total fidelity of our ability to prepare and readout the defect's charge state in a single shot. Despite the absence of solid-immersion lenses and other photonically enhancing structures, we achieve a high number of photons per shot. The non-unity fidelity likely arises from imperfect charge state preparation due to optical excitation of nearby traps, as we discuss in the following sections. The high single-shot photon number and the near-unity fidelity of the charge readout technique exemplify its advantage over traditional spin-photon readout.

Fig. 2C shows the projected number of photons per-shot during these charge readout windows using various resonant laser powers. With increased laser power, more photons are scattered per second, but additional two-photon ionization occurs, reducing the time during which the charge state can be read. On the other hand, if the laser power is too low, very few photons per second are scattered but readout time is still limited by the charge lifetime and the readout window cannot be arbitrarily extended. As a result, there is an optimal choice of laser power to maximize the readout, which in our case results in over 1500 photons per-shot at ~1 µW. This behavior is understood with a simple predictive model (Supplementary Materials). The high number of scattered photons means that single-shot readout of the charge state is possible even with extremely low collection efficiency or in systems with much lower quantum yield.

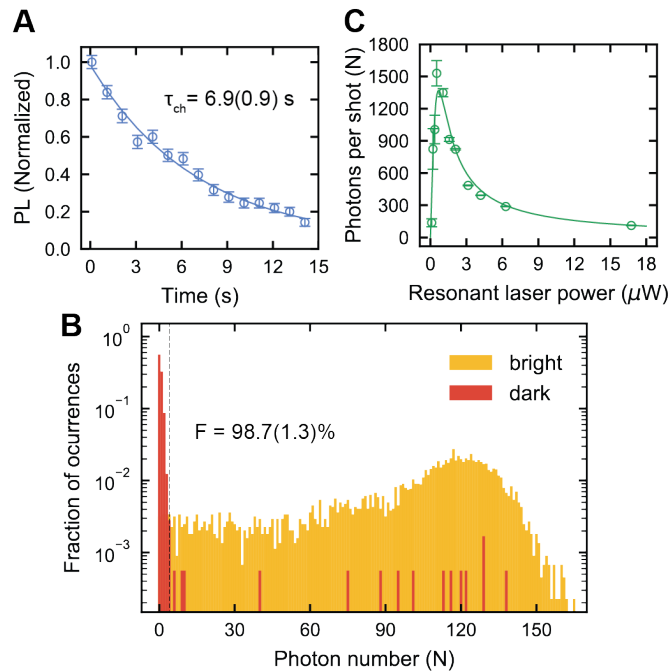

**Fig. 2. Single-shot readout of the divacancy charge state.** (**A**) Charge readout PL signal dependence on delay time between the charge initialization and readout follows an exponential decay $e^{\left(-\frac{t}{\tau_{ch}}\right)}$ where $\tau_{ch}$ is the charge lifetime. We find that $\tau_{ch}$ is 6.9(0.9) seconds. (**B**) Log-scale histogram of photon number distributions collected during a charge readout for preparation into the neutral 'bright' state and ionized 'dark' state. We use a 20 ms readout window with 4.05 µW combined resonant laser power, selected to maximize the readout fidelity. For a cutoff of $N=4$ photons, the single-shot readout fidelity of the charge state is 98.7(1.3)%. The false positive rate $p_{0|1} = 1.17\%$ and false negative rate $p_{1|0} = 1.26\%$. (**C**) Extracted photons per shot from observed PL rate and charge state decay for various combined resonant laser powers. The maximal extracted photons per shot is $N=1529(117)$. The line is a fit from a model (Supplementary Materials). All data is taken at B=18 G and T=5 K. All reported errors represent 1 SE from the fit and all error bars represent 1 SD of the raw data.

Having established our ability to perform high-fidelity single-shot readout of the charge state, we next characterize the various charge transition processes induced by the lasers used in our experiments. We first characterize the rate at which we initialize the divacancy to the neutral state with the 705 nm charge repump laser. After preparing the defect in the dark charge state, a time-varying charge initialization pulse is applied to reset the defect to the neutral, bright state. Using a charge readout, we measure the recovery rate of the bright state as a function of the 705 nm laser power (Fig. 3A). The charge repump rate is 993±17 MHz/W, consistent with the one-photon repumping rate described in (*11*).

We next characterize the ionization rate solely from the resonant lasers and ionization laser, respectively, by initializing into the bright state and measuring the charge state after a variable length laser pulse. The exponentially-fit ionization rates for various powers of the resonant lasers are displayed in Fig. 3B. The ionization rate is initially quadratic and then increases linearly as 10.6±0.9 MHz/W, signaling saturation of the optical transition (Supplementary Materials). At the relevant resonant laser powers in our experiment, we observe ionization rates around 100 Hz. On the other hand, ionization from solely the 1151 nm laser increases linearly with power as 95.7±3.7 kHz/W (Fig. 3C). Ideally, the 1151 nm laser by itself should cause no ionization, where here a small residual rate likely arises from excitation of nearby traps, freeing carriers that alter the defect's charge state.

Finally, we characterize our ability to ionize the defect once it is in its optical excited state, which is the prerequisite for spin-dependent ionization. After charge initializing to the bright state, we spin-agnostically ionize the defect using a variable length pulse where both the $E_x$ and $E_{1,2}$ resonant lasers and the 1151 nm ionization laser are on at the same time. For these experiments, the resonant power is kept such that the defect optical transition is saturated. The decay rate of the signal from charge readout is displayed for various ionization powers in Fig. 3D, where the saturating behavior can be understood through the effect of stimulated emission, discussed later (Supplementary Materials).

Importantly, the spin-agnostic ionization rates (order MHz) are nearly three orders of magnitude greater than the unwanted ionization rate from only the 1151 nm laser (order kHz) or only the resonant lasers (order 100 Hz). This confirms that for experiments where both resonant excitation and the 1151 nm ionization laser are used, the dominant source of ionization is a two-photon process where one photon induces a resonant, ground to excited state transition of the defect and a second photon from the ionization laser subsequently converts the defect to its negative charge state.

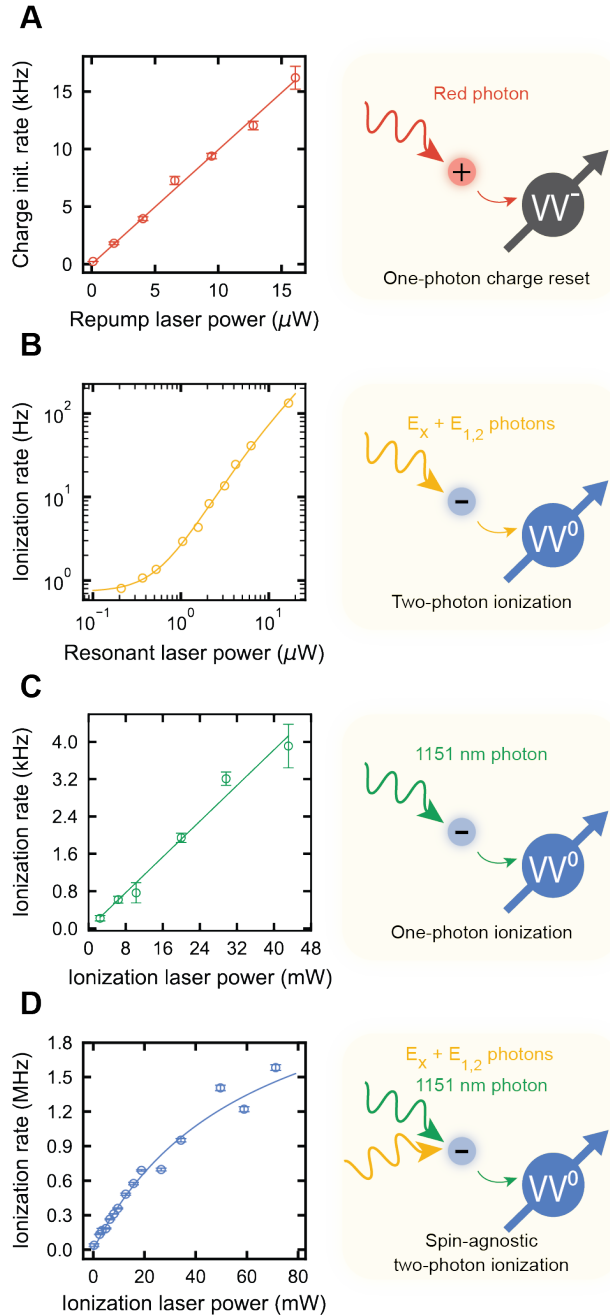

**Fig. 3 Optical charge reset and ionization processes.** (**A**) Power dependence of the charge reset rate using the 705 nm laser. The reset rate is linear (red line fit) with power as 993±17 Hz/µW (**B**) Ionization rate dependence on combined resonant laser power ($E_x$ and $E_{1,2}$ lines). The solid line is a fit using a saturating two-photon ionization model (Supplementary Materials). (**C**) Ionization rate dependence on 1151 nm laser alone. The ionization rate is linear with power (solid line fit) as 95.7±3.7 kHz/W. (**D**) Spin-agnostic ionization rate dependence on the 1151 nm laser ionization power. The resonant laser excitation is beyond saturation at 15 µW. The solid blue line is a fit from our model including the effect of stimulated emission (Supplementary Materials). All data is taken at B= 18 G and T= 5 K. All reported errors represent 1 SE from the fit and all error bars represent 1 SD of the raw data.

**Spin-to-charge control**

Given that we can perform high-fidelity, single-shot readout of the divacancy charge state, and that we can ionize the defect with a combination of resonant and red-detuned laser light, we can selectively map the divacancy spin state onto its charge state to achieve single-shot readout of the spin state. Specifically, after charge initialization to VV$^0$ and spin initialization to $m_s=0$, we spin-selectively photo-ionize the $m_s=0$ state to VV$^-$ by simultaneous excitation of the E$_x$ optical transition ($m_s=0$ character) while applying the 1151 nm ionization laser. The defect can be protected from this spin-selective ionization by rotating into $m_s=-1$ via the application of a microwave $\pi$ pulse so that the laser no longer optically excites it. Thus, a spin initialized to $m_s=-1$ does not undergo ionization and remains in VV$^0$, forming the basis of spin contrast for the SCC process. Importantly, SCC is performed using the E$_x$ transition due to its high cyclicity, which increases the number of times the excited state can be populated before a spin-flip occurs. Spin-flips cause destruction of the spin state (*33*) and prevent ionization from occurring, therefore reducing the fidelity of the conversion process (Supplementary Materials). Thus, a key aspect of the SCC process is ensuring the rate of spin-selective ionization exceeds the rate of spin-flip errors.

After SCC, we perform single-shot readout of the charge state. The charge readout signal for states prepared to $m_s=0$ and $m_s=-1$ as the SCC pulse duration is swept is shown in Fig. 4A. After the SCC contrast reaches a maximum, the contrast decreases with increased spin-selective ionization pulse durations due to a reduction in PL from states prepared into $m_s=-1$. This decreasing PL is caused by non-spin selective ionization, but has a rate of decay that exceeds the ionization rate from only the 1151 nm laser (Fig. 3C). Therefore, we attribute this non-spin selective ionization to other mechanisms such as weak excitation of the $m_s=\pm 1$ optical transitions by the E$_x$ laser, or heating effects induced by the high-power ionization laser that cause additional orbital or spin mixing.

We next characterize the end-to-end fidelity of the combined initialization, SCC, and readout process by examining the resulting single-shot photon number distribution. Fig. 3B shows histograms of the photon statistics for states prepared into $m_s=0$ and $m_s=-1$. We extract a maximum end-to-end SCC fidelity of $F_{SCC}=80.8\pm0.6\%$ (Supplementary Materials). When corrected for the charge initialization and charge readout fidelity, we obtain a SCC conversion fidelity of $F_{SCC}'=F_{SCC}/F_{charge}=81.6\%$, revealing that the main source of infidelity is the SCC conversion process and not the charge readout. We eliminate infidelity arising from errors in spin manipulation and selectivity of the spin-photon interface, due to our observation of over 99% Rabi contrast in fluorescence readout, consistent with previous reports (*7*). This fidelity is greater than the SCC contrast (Fig. 4A) due to non-zero background counts and appreciable ionization during the readout window. We track the SCC end-to-end fidelity ($F_{SCC}$) while sweeping the SCC pulse duration (Fig. 4C) and find that fidelity is maximized for a pulse duration of approximately ~2 μs. We note that in Fig. 4B, there is significant population distribution above and below the $N=2$ single-shot cutoff for both spin preparations, corresponding to a false positive rate $p_{0|1}=27\%$ and false negative rate $p_{1|0}=11\%$, for $m_s=0$ and $m_s=-1$, respectively (Supplementary Materials). This indicates that incomplete ionization of $m_s=0$ is the dominant source of infidelity in our SCC process, which we discuss below.

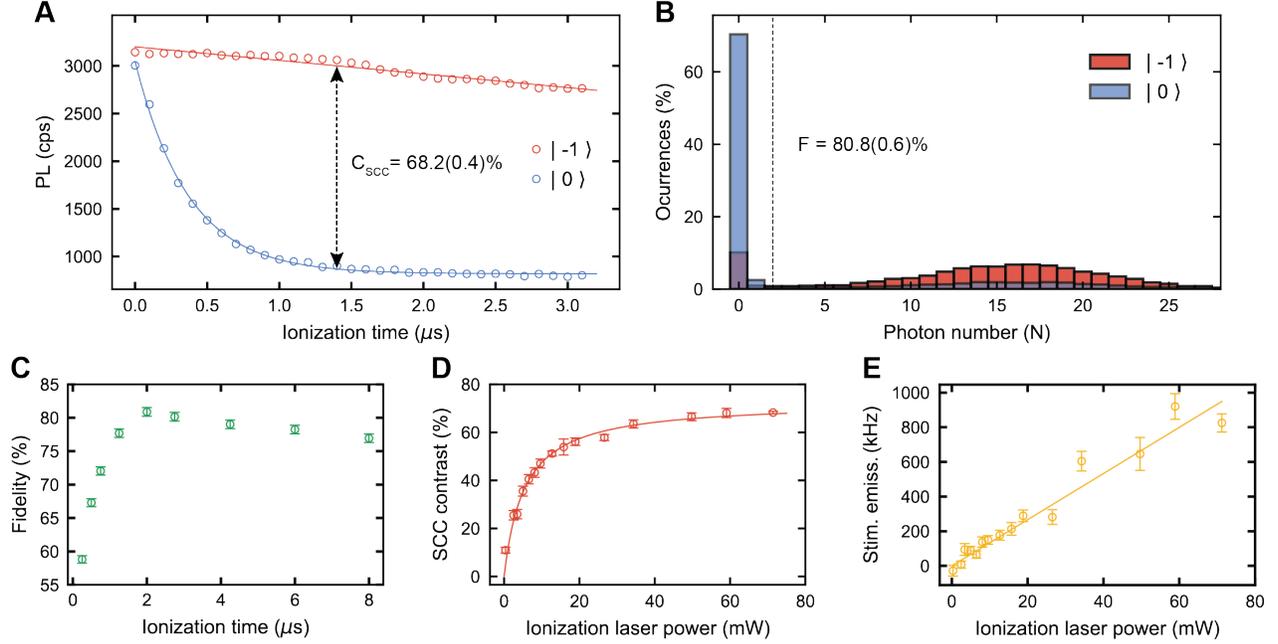

**Fig. 4. Single-shot readout of the spin state with spin-to-charge conversion.** (**A**) Charge readout signal following SCC step for preparation into $m_s = 0$ and $m_s = -1$. The maximum fitted contrast is 68.2(0.4)% at an ionization laser power of 71 mW and a SCC pulse duration of $t_{ion}$ = 1.39 µs, using 14.95 µW of resonant power. (**B**) Charge readout photon number distribution after SCC step for preparation into $m_s = 0$ and $m_s = -1$. The end-to-end process fidelity is 80.8(0.6)% for a cutoff of $N=2$ photons. (**C**) Dependence of SCC fidelity on SCC pulse duration. (**D**) Dependence of SCC contrast with 1151 nm ionization laser power. The contrast follows a saturation behavior (red line fit, Supplementary Materials). (**E**) Calculated stimulated emission rate dependence on 1151 nm ionization laser power. The stimulated emission rate increases linearly as 13.3 MHz/W (yellow line fit). All data is taken at B=18 G and T=5 K. All reported errors represent 1 SE from the fit and all error bars represent 1 SD of the raw data.

Although we increase the ionization laser power to maximize SCC contrast and fidelity, we observe that the SCC contrast ($C_{SCC}$) unexpectedly saturates with power (Fig. 4D), limiting our single-shot readout fidelity. We attribute this saturation behavior to stimulated emission from the excited state induced by the 1151 nm ionization laser (*39*). Stimulated emission induces an excited to ground state transition of the defect, effectively increasing the spin-flip rate and decreasing the occupation time in the excited state. This results in a reduced chance of ionization via SCC before a spin-flip occurs. We model the dynamics of the SCC process using a set of differential equations to describe the various rates of ionization and spin-flips in our system (Supplementary Materials). Using the measured spin-agnostic ionization rate (Fig. 3D), the spin-flip rate, and the time evolution of Fig. 4A, our model predicts a maximum SCC fidelity of about 72%, which is consistent with our experimental findings. We also find that stimulated emission results in the saturation behavior of Fig. 3D and Fig. 4D, and are able to extract the stimulated emission rate in Fig. 4E (Supplementary Materials).

Importantly, we find that the simple ratio of the ionization cross section from the excited state to the stimulated emission cross section ($\sigma_i/\sigma_s$) determines the resulting fidelity of SCC, where larger ratios are desirable. We directly calculate this metric with DFT, which indicates that the ratio is only optimal for a narrow energy window below the ZPL energy and above the energy of the first vibronic peak in the emission sideband (Fig. S4B, S4D). Our theoretical results also show that the cross sections and the ratio $\sigma_i/\sigma_s$ do not change as a function of the light polarization when the incident light is parallel to the 3-fold rotation axis ($C_{3v}$) of the defect (Fig S5A, S5B). Surprisingly, however, a significant increase in the ratio $\sigma_i/\sigma_s$ may be obtained by using polarized light perpendicular to the defect axis (Supplementary Materials). This suggests that a change in the ionization laser geometry and polarization may drastically increase the SCC fidelity. These investigations into the limitations of SCC due to stimulated emission through DFT and modelling create a set of guidelines and considerations for designing qubits and optimizing these types of spin-to-charge experiments.

### Extending coherence with dynamical decoupling

Having demonstrated the ability to spin-selectively ionize the defect, we take advantage of the single-shot readout afforded by the SCC technique and perform measurements that reveal the exceptionally long spin coherence time of the defect's spin state. We first perform $T_1$ relaxation measurements using SCC readout (Fig. 5A). Although the charge state has appreciable decay on these timescales, we normalize our measurements such that we can extract the pure spin relaxation time. Despite no obvious spin $T_1$ decay in Fig. 5A, we investigate the $T_1$ time using a $\chi^2$ square goodness of fit test and place a lower bound of 103 seconds on the $T_1$ time with 95% confidence (Supplementary Materials). This minute-scale lower bound eliminates $T_1$ spin relaxation as a concern for this system, and is on par with the longest reported times for the NV$^-$ center in diamond (*27*)

In natural SiC, spin decoherence is dominated by magnetic fluctuations from flip-flop interactions between $^{29}$Si and $^{13}$C possessing I=½ nuclear spin (*1*). The sample studied here was isotopically engineered to reduce the occurrence of nuclear flip-flops and extend the coherence time (*7*) (Materials and Methods). In this work, we further extend this spin coherence by applying dynamical decoupling sequences (*23*). Fig. 5B displays the coherence for sequences consisting of $N = 1$ to $N=16384$ pulses, measured with single-shot readout. With the combination of isotopic purification and dynamical decoupling, we measure a maximum $T_2$ time of 5.3±1.3 seconds, an improvement of over two orders of magnitude over the previously reported extended coherences in the divacancy system (*2*).

The dependence of coherence ($T_2$) with pulse number ($N$) (Fig. 5C) can be modeled as $T_2 \sim N^\psi$ where $\psi$ varies with the frequency cutoff, shape, and roll-off behavior of sources of dephasing in the system (*40*). We find that the scaling of coherence for low pulse numbers has $\psi=0.92\pm0.01$ and at high pulse numbers $\psi=0.75\pm0.01$ (Fig. 5C). This may suggest that dephasing is dominated by two separate noise sources, each with differing associated frequency cutoffs that dominate in different regimes (*41*). The existence of these two competing noise sources is supported by previous work on the same isotopically purified sample that suggests both a fast paramagnetic and slow nuclear spin bath play a role in dephasing (*7*), though certain broad Lorentzian baths may exhibit a similar $\psi \sim 1$ (*40*) to $\psi \sim 2/3$ crossover (*28*). Another possibility is that our experiments become dominated by control errors at high pulse number and do not protect the state as effectively, as we observe the contrast decreasing in this regime.

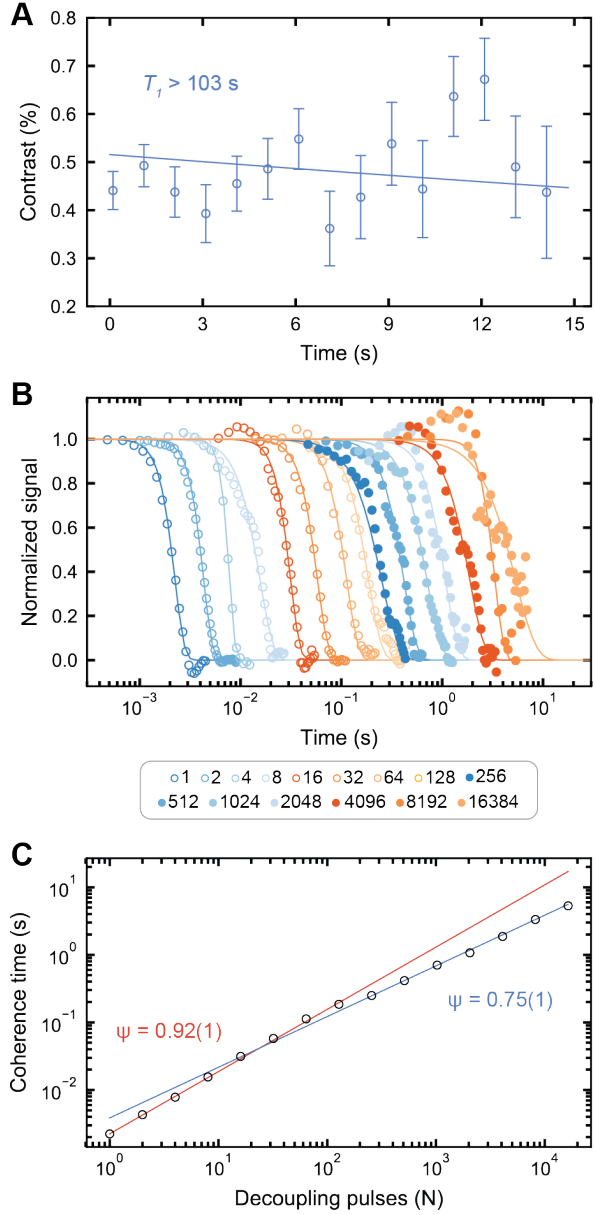

**Fig. 5. Ultralong spin coherence and lifetime for a single divacancy.** (**A**) $T_1$ spin relaxation time of divacancy using single-shot readout. Using a goodness of fit test with a 95% confidence interval, we estimate that $T_1 \geq 103$ seconds (Supplementary Materials). The fit (solid line) is for a $T_1$ of 103 seconds. (**B**) $T_2$ decay curves measured after applying dynamical decoupling pulses sequences of increasing pulse number, $N$. We avoid ESEEM oscillations by enforcing pulse spacing requirements as in Ref. (*27*), eliminating sharp dips, and smoothing to find the coherence function envelope. The envelope is fit to a stretched exponential function $Ae^{-\left(\frac{t}{\tau}\right)^n}$ where $n$ is a stretch factor. (**C**) Extension of $T_2$ coherence time with total decoupling pulse number. We fit in log space the low (blue) and high (red) pulse number regimes as $T_2 \sim N^\psi$. All data is taken at B=18 G and T=5 K. All reported errors represent 1 SE from the fit and all error bars represent 1 SD of the raw data.

Even after over $10^4$ pulses, the measured coherence time does not saturate. Given the long $T_1$ times, we expect coherence times greater than five seconds to be possible with even longer decoupling sequences (Materials and Methods). The combination of dynamical decoupling and isotopic purification in this work results in coherences that exceed state-of-the-art single spin measurements in other competing systems (*27*). These ultralong coherences offer many advantages for SiC-based quantum technologies. For example, in ac sensing protocols, long coherence times extend the phase accumulation period and increase sensitivity to weak signals. Additionally, the reduction of memory storage errors that come with long coherence times is vital for the development of quantum repeaters, which are a necessary component for future quantum networks.

**Discussion**
The all-optical spin-to-charge conversion technique demonstrated here can be extended to other emitters where single-shot readout is needed (*42–44*), particularly in platforms where the photonic devices typically used to boost collection efficiency can degrade charge and optical qualities, or where detector or emitter quantum efficiency is low. Additionally, coherence protection combined with the high-fidelity readout provided by the SCC technique brings the SiC material platform to the forefront for use in sensing and communication protocols that require both deterministic readout and quantum state preservation.

Looking forward, use of auxiliary microwave drives (*19*) to recover leakage of population from the qubit states, optimization of the ionization laser wavelength to maximize $\sigma_i/\sigma_s$, and further extension of the charge state stability via growth techniques may present pathways to improving the SCC technique. Our DFT calculations additionally suggest that a significant reduction in undesirable stimulated emission can be achieved by employing polarized light with incidence orthogonal to the defect axis, while still effectively ionizing the defect. We note, however, that the stimulated emission from defects is also of fundamental interest for future developments of lasers and new kinds of sensors (*45*). Furthermore, the charge itself may lead to new sensing opportunities (*46*), while the use of electrical depletion can reduce some of the remaining noise in the system to further increase coherence (*11*, *47*, *48*).

Integration of SCC with the optical tuning capabilities offered by SiC devices, such as p-i-n diodes (*11*), could be used to construct highly scalable, tunable SiC quantum nodes for future entanglement schemes. More generally, the SCC technique unlocks a new generation of quantum devices with single charge sensitivity, where divacancy-embedded classical semiconductor structures such as MOSFETs, diodes, and APDs could be used to bridge the gap between quantum and classical electronic devices.

**Materials and Methods**
  **Experimental details**

Measurements are performed at T = 5 K in closed-cycle Montana Cryostat with a 0.85 NA near-infrared objective. All measurements are performed at low field of around ~18 Gauss on a single *hh* divacancy in 4H-SiC. We use two Toptica DLC PRO lasers for narrow-line laser control of the defects. The 1151 nm ionization laser is a QPhotonics QFLD-1160-300S temperature tunable laser diode. PL is detected with a Quantum Opus superconducting nanowire single photon detector. Photon counting and time tagging experiments use a PicoQuant Picoharp 300 Time-Correlated Single Photon Counting (TCSPC) system. Spin initialization is achieved by pumping on the $E_{1,2}$ transitions, while the $E_x$ transition serves as the cycling transition for readout and spin-to-charge conversion. When performing SCC, to mitigate the effects of this non-selective ionization, we saturate the defect with resonant laser excitation to maximize the chances of desirable spin-selective ionization during the SCC step before a charge conversion error occurs. Additionally, for subsequent experiments we utilize the highest possible ionization laser power of 71 mW to increase the likelihood of ionization during the SCC step. Driving of transitions between the $m_s = 0$ and $m_s = \pm 1$ spin sublevels is performed using 1.357 GHz microwaves applied through Ti/Au striplines patterned on the sample surface. Microwave extinction and filtering is an important parameter to increase coherence and lifetime. We use a 800 MHz highpass filter after amplification, and switch the microwave pulses on and off with a ZASWA-2-50DR+ both before and after amplification. Further extinction is achieved utilizing the IQ modulation of the source, which also provides phase control of the microwave pulses. We use this phase control to perform the XY-8 pulse sequences for dynamical decoupling which reduces sensitivity to pulse imperfections and drift. For all spin control experiments, we use π pulses with a length of approximately 1-2 μs. We stop our experiments at pulse number *N*=16384 for CPMG due to compounding pulse infidelity that reduces contrast and because our measured coherence approaches the charge lifetime for this defect. In Fig. 5B, the coherence function is normalized such that the contrast reduction and the finite charge lifetime do not affect the fitted coherences. For the coherence measurements, two fiber-coupled AOMs (AA Opto-electronic MT250-IR6-Fio-SMO) are used in series on the resonant lasers to achieve high extinction. Pulsing of the red charge reset tone (705 nm) and the 1151 nm ionization laser is achieved with direct modulation of the laser diode with a Thorlabs CLD1015 diode control unit. Extinction from this modulation is high, where we note the finite charge lifetime is not affected by extra extinction of the 705 nm laser in the off state. All parameters in the text have errors reported at one standard error (SE).

  **Isotopically purified sample**

The sample consists of epitaxial 4H-SiC grown by chemical vapor deposition on a 4° off-axis n-type 4H-SiC. The layer thickness is ~90 μm, and uses isotopically purified Si and C precursor gasses as in reference (*7*). Secondary ion mass spectroscopy reveals purities of 99.85% $^{28}$Si and 99.98% $^{12}$C. C-V measurements show slightly n-type behavior with a carrier concentration of $6 \times 10^{13}$ cm$^{-3}$. Single defects are created using a $1 \times 10^{13}$ cm$^{-2}$ dose of 2-MeV relativistic electrons. Subsequent annealing at 810 °C in Ar gas results in isolated single VV$^0$. We note that this slight n-type behavior causes VV$^0$ to be unstable under illumination, where the negative charge states are favored. This is key, however, to our ability to ionize the VV$^0$ effectively. The available carriers from these dopants provide the necessary charges to continually source and redistribute charges for these processes. SCC naturally requires charge unstable defects.

# Computational details

We carried out hybrid DFT calculations to determine the excitation energies of the VV$^0$ defect and the charge transition energy to the VV$^-$ defect. All calculations were performed using the dielectric-dependent hybrid (DDH) functional (*49*) and the Quantum Espresso code (*50*). We used a 5x5x2 Γ-centered supercell, SG15 ONCV pseudopotentials (*51*) and a plane-wave basis set with a kinetic energy cut-off of 80 Ry. In the case of charged defects, we applied corrections to the total energy as derived in (*52*). We computed the optical matrix elements pertaining to the ionization and stimulated emission cross sections using a Γ-centered supercell with 1296 atomic sites and the PBE functional (*53*). Electron-phonon spectral functions were computed using the generating function approach within the displaced harmonic approximation (*39, 54, 55*). The phonon modes of the defective solid were obtained using a 5x5x2 supercell generated with the PHONONPY (*56*) package and extrapolated to larger sizes (16x16x5 supercell). Additional details are reported in the Supplementary Materials.

## References

1. H. Seo, A. L. Falk, P. v Klimov, K. C. Miao, G. Galli, D. D. Awschalom, Quantum decoherence dynamics of divacancy spins in silicon carbide. *Nature Communications*. **7**, 12935 (2016).
2. K. C. Miao, J. P. Blanton, C. P. Anderson, A. Bourassa, A. L. Crook, G. Wolfowicz, H. Abe, T. Ohshima, D. D. Awschalom, Universal coherence protection in a solid-state spin qubit. *Science*. **369**, 1493–1497 (2020).
3. D. J. Christle, A. L. Falk, P. Andrich, P. v. Klimov, J. U. Hassan, N. T. Son, E. Janzén, T. Ohshima, D. D. Awschalom, Isolated electron spins in silicon carbide with millisecond coherence times. *Nature Materials*. **14**, 160–163 (2015).
4. S. Kanai, F. J. Heremans, H. Seo, G. Wolfowicz, C. P. Anderson, S. E. Sullivan, G. Galli, D. D. Awschalom, H. Ohno, Generalized scaling of spin qubit coherence in over 12,000 host materials (2021) (available at http://arxiv.org/abs/2102.02986).
5. D. J. Christle, P. v. Klimov, C. F. de las Casas, K. Szász, V. Ivády, V. Jokubavicius, J. U. Hassan, M. Syväjärvi, W. F. Koehl, T. Ohshima, N. T. Son, E. Janzén, ádám Gali, D. D. Awschalom, Isolated spin qubits in SiC with a high-fidelity infrared spin-to-photon interface. *Physical Review X*. **7**, 1–12 (2017).
6. K. C. Miao, A. Bourassa, C. P. Anderson, S. J. Whiteley, A. L. Crook, S. L. Bayliss, G. Wolfowicz, G. Thiering, P. Udvarhelyi, V. Ivády, H. Abe, T. Ohshima, Á. Gali, D. D. Awschalom, Electrically driven optical interferometry with spins in silicon carbide. *Science Advances*. **5**, eaay0527 (2019).
7. A. Bourassa, C. P. Anderson, K. C. Miao, M. Onizhuk, H. Ma, A. L. Crook, H. Abe, J. Ul-Hassan, T. Ohshima, N. T. Son, G. Galli, D. D. Awschalom, Entanglement and control of single nuclear spins in isotopically engineered silicon carbide. *Nature Materials*. **19**, 1319–1325 (2020).
8. G. Wolfowicz, F. J. Heremans, C. P. Anderson, S. Kanai, H. Seo, A. Gali, G. Galli, D. D. Awschalom, Quantum guidelines for solid-state spin defects. *Nature Reviews Materials* (2021), doi:10.1038/s41578-021-00306-y.
9. A. L. Crook, C. P. Anderson, K. C. Miao, A. Bourassa, H. Lee, S. L. Bayliss, D. O. Bracher, X. Zhang, H. Abe, T. Ohshima, E. L. Hu, D. D. Awschalom, Purcell Enhancement of a Single Silicon Carbide Color Center with Coherent Spin Control. *Nano Letters*. **20**, 3427–3434 (2020).
10. D. M. Lukin, C. Dory, M. A. Guidry, K. Y. Yang, S. D. Mishra, R. Trivedi, M. Radulaski, S. Sun, D. Vercruysse, G. H. Ahn, J. Vučković, 4H-silicon-carbide-on-insulator for integrated quantum and nonlinear photonics. *Nature Photonics*. **14**, 330–334 (2020).
11. C. P. Anderson, A. Bourassa, K. C. Miao, G. Wolfowicz, P. J. Mintun, A. L. Crook, H. Abe, J. U. Hassan, N. T. Son, T. Ohshima, D. D. Awschalom, Electrical and optical control of single spins integrated in scalable semiconductor devices. *Science*. **366**, 1225–1230 (2019).
12. S. J. Whiteley, G. Wolfowicz, C. P. Anderson, A. Bourassa, H. Ma, M. Ye, G. Koolstra, K. J. Satzinger, M. v. Holt, F. J. Heremans, A. N. Cleland, D. I. Schuster, G. Galli, D. D. Awschalom, Spin–phonon interactions in silicon carbide addressed by Gaussian acoustics. *Nature Physics*. **15**, 490–495 (2019).
13. S. J. Whiteley, F. J. Heremans, G. Wolfowicz, D. D. Awschalom, M. v. Holt, Correlating dynamic strain and photoluminescence of solid-state defects with stroboscopic x-ray diffraction microscopy. *Nature Communications*. **10**, 5–10 (2019).

14. L. Robledo, L. Childress, H. Bernien, B. Hensen, P. F. A. Alkemade, R. Hanson, High-fidelity projective read-out of a solid-state spin quantum register. *Nature*. **477**, 574–578 (2011).
15. N. T. Son, C. P. Anderson, A. Bourassa, K. C. Miao, C. Babin, M. Widmann, M. Niethammer, J. Ul Hassan, N. Morioka, I. G. Ivanov, F. Kaiser, J. Wrachtrup, D. D. Awschalom, Developing silicon carbide for quantum spintronics. *Applied Physics Letters*. **116**, 190501 (2020).
16. H. Bernien, B. Hensen, W. Pfaff, G. Koolstra, M. S. Blok, L. Robledo, T. H. Taminiau, M. Markham, D. J. Twitchen, L. Childress, R. Hanson, Heralded entanglement between solid-state qubits separated by three metres. *Nature*. **497**, 86–90 (2013).
17. J. Cramer, N. Kalb, M. A. Rol, B. Hensen, M. S. Blok, M. Markham, D. J. Twitchen, R. Hanson, T. H. Taminiau, Repeated quantum error correction on a continuously encoded qubit by real-time feedback. *Nature Communications*. **7**, 1–7 (2016).
18. D. M. Irber, F. Poggiali, F. Kong, M. Kieschnick, T. Lühmann, D. Kwiatkowski, J. Meijer, J. Du, F. Shi, F. Reinhard, Robust all-optical single-shot readout of nitrogen-vacancy centers in diamond. *Nature Communications*. **12** (2021), doi:10.1038/s41467-020-20755-3.
19. Q. Zhang, Y. Guo, W. Ji, M. Wang, J. Yin, F. Kong, Y. Lin, C. Yin, F. Shi, Y. Wang, J. Du, High-fidelity single-shot readout of single electron spin in diamond with spin-to-charge conversion. *Nature Communications*. **12**, 1–6 (2021).
20. J. R. Maze, P. L. Stanwix, J. S. Hodges, S. Hong, J. M. Taylor, P. Cappellaro, L. Jiang, M. V. G. Dutt, E. Togan, A. S. Zibrov, A. Yacoby, R. L. Walsworth, M. D. Lukin, Nanoscale magnetic sensing with an individual electronic spin in diamond. *Nature*. **455**, 644–647 (2008).
21. J. M. Taylor, P. Cappellaro, L. Childress, L. Jiang, D. Budker, P. R. Hemmer, A. Yacoby, R. Walsworth, M. D. Lukin, High-sensitivity diamond magnetometer with nanoscale resolution. *Nature Physics*. **4**, 810–816 (2008).
22. C. L. Degen, F. Reinhard, P. Cappellaro, Quantum sensing. *Reviews of Modern Physics*. **89**, 1–39 (2017).
23. T. Gullion, D. B. Baker, M. S. Conradi, New, compensated Carr-Purcell sequences. *Journal of Magnetic Resonance (1969)*. **89**, 479–484 (1990).
24. H. Y. Carr, E. M. Purcell, Effects of Diffusion on Free Precession in Nuclear Magnetic Resonance Experiments. *Phys. Rev.* **94**, 630–638 (1954).
25. J. T. Muhonen, J. P. Dehollain, A. Laucht, F. E. Hudson, R. Kalra, T. Sekiguchi, K. M. Itoh, D. N. Jamieson, J. C. McCallum, A. S. Dzurak, A. Morello, Storing quantum information for 30 seconds in a nanoelectronic device. *Nature Nanotechnology*. **9**, 986–991 (2014).
26. A. M. Tyryshkin, S. Tojo, J. J. L. Morton, H. Riemann, N. v Abrosimov, P. Becker, H.-J. Pohl, T. Schenkel, M. L. W. Thewalt, K. M. Itoh, S. A. Lyon, Electron spin coherence exceeding seconds in high-purity silicon. *Nature Materials*. **11**, 143–147 (2012).
27. M. H. Abobeih, J. Cramer, M. A. Bakker, N. Kalb, M. Markham, D. J. Twitchen, T. H. Taminiau, One-second coherence for a single electron spin coupled to a multi-qubit nuclear-spin environment. *Nature Communications*. **9**, 1–8 (2018).
28. N. Bar-Gill, L. M. Pham, A. Jarmola, D. Budker, R. L. Walsworth, Solid-state electronic spin coherence time approaching one second. *Nature Communications*. **4** (2013), doi:10.1038/ncomms2771.
29. D. Simin, H. Kraus, A. Sperlich, T. Ohshima, G. v. Astakhov, V. Dyakonov, Locking of electron spin coherence above 20 ms in natural silicon carbide. *Physical Review B*. **95**, 161201 (2017).
30. N. T. Son, P. Carlsson, J. ul Hassan, E. Janzén, T. Umeda, J. Isoya, A. Gali, M. Bockstedte, N. Morishita, T. Ohshima, H. Itoh, Divacancy in 4H-SiC. *Physical Review Letters*. **96**, 055501 (2006).
31. P. Tamarat, N. B. Manson, J. P. Harrison, R. L. McMurtrie, A. Nizovtsev, C. Santori, R. G. Beausoleil, P. Neumann, T. Gaebel, F. Jelezko, P. Hemmer, J. Wrachtrup, Spin-flip and spin-conserving optical transitions of the nitrogen-vacancy centre in diamond. *New Journal of Physics*. **10** (2008), doi:10.1088/1367-2630/10/4/045004.
32. N. B. Manson, J. P. Harrison, M. J. Sellars, Nitrogen-vacancy center in diamond: Model of the electronic structure and associated dynamics. *Physical Review B - Condensed Matter and Materials Physics*. **74**, 1–11 (2006).
33. L. Gordon, A. Janotti, C. G. van de Walle, Defects as qubits in 3C- and 4H-SiC. *Physical Review B - Condensed Matter and Materials Physics*. **92**, 1–5 (2015).
34. G. Wolfowicz, C. P. Anderson, A. L. Yeats, S. J. Whiteley, J. Niklas, O. G. Poluektov, F. J. Heremans, D. D. Awschalom, Optical charge state control of spin defects in 4H-SiC. *Nature Communications*. **8**, 1876 (2017).

35. D. A. Golter, C. W. Lai, Optical switching of defect charge states in 4H-SiC. *Scientific Reports*. **7**, 1–5 (2017).
36. B. Magnusson, N. T. Son, A. Csóré, A. Gällström, T. Ohshima, A. Gali, I. G. Ivanov, Excitation properties of the divacancy in $4H$-SiC. *Phys. Rev. B*. **98**, 195202 (2018).
37. M. Bockstedte, F. Schütz, T. Garratt, V. Ivády, A. Gali, Ab initio description of highly correlated states in defects for realizing quantum bits. *npj Quantum Materials*. **3**, 31 (2018).
38. D. A. Hopper, H. J. Shulevitz, L. C. Bassett, Spin Readout Techniques of the Nitrogen-Vacancy Center in Diamond. *Micromachines*. **9** (2018), doi:10.3390/mi9090437.
39. L. Razinkovas, M. Maciaszek, F. Reinhard, M. W. Doherty, A. Alkauskas, Photoionization of negatively charged NV centers in diamond: theory and ab initio calculations (2021) (available at http://arxiv.org/abs/2104.09144).
40. M. J. Biercuk, H. Bluhm, Phenomenological study of decoherence in solid-state spin qubits due to nuclear spin diffusion. *Physical Review B*. **83**, 235316 (2011).
41. J. Medford, Ł. Cywiński, C. Barthel, C. M. Marcus, M. P. Hanson, A. C. Gossard, Scaling of Dynamical Decoupling for Spin Qubits. *Physical Review Letters*. **108**, 086802 (2012).
42. G. Wolfowicz, C. P. Anderson, B. Diler, O. G. Poluektov, F. J. Heremans, D. D. Awschalom, Vanadium spin qubits as telecom quantum emitters in silicon carbide. *Science Advances*. **6**, eaaz1192 (2020).
43. B. Diler, S. J. Whiteley, C. P. Anderson, G. Wolfowicz, M. E. Wesson, E. S. Bielejec, F. Joseph Heremans, D. D. Awschalom, Coherent control and high-fidelity readout of chromium ions in commercial silicon carbide. *npj Quantum Information*. **6**, 11 (2020).
44. R. Nagy, M. Niethammer, M. Widmann, Y.-C. Chen, P. Udvarhelyi, C. Bonato, J. U. Hassan, R. Karhu, I. G. Ivanov, N. T. Son, J. R. Maze, T. Ohshima, Ö. O. Soykal, Á. Gali, S.-Y. Lee, F. Kaiser, J. Wrachtrup, High-fidelity spin and optical control of single silicon-vacancy centres in silicon carbide. *Nature Communications*. **10**, 1954 (2019).
45. J. Jeske, D. W. M. Lau, X. Vidal, L. P. McGuinness, P. Reineck, B. C. Johnson, M. W. Doherty, J. C. McCallum, S. Onoda, F. Jelezko, T. Ohshima, T. Volz, J. H. Cole, B. C. Gibson, A. D. Greentree, Stimulated emission from nitrogen-vacancy centres in diamond. *Nature Communications*. **8**, 14000 (2017).
46. G. Wolfowicz, C. P. Anderson, S. J. Whiteley, D. D. Awschalom, Heterodyne detection of radio-frequency electric fields using point defects in silicon carbide. *Applied Physics Letters*. **115** (2019), doi:10.1063/1.5108913.
47. D. R. Candido, M. E. Flatté, Suppression of the optical linewidth and spin decoherence of a quantum spin center in a *p-n* diode (2020) (available at https://arxiv.org/abs/2008.13289).
48. K. Ghosh, H. Ma, M. Onizhuk, V. Gavini, G. Galli, Spin–spin interactions in defects in solids from mixed all-electron and pseudopotential first-principles calculations. *npj Computational Materials*. **7**, 123 (2021).
49. J. H. Skone, M. Govoni, G. Galli, Self-consistent hybrid functional for condensed systems. *Physical Review B*. **89**, 195112 (2014).
50. P. Giannozzi, S. Baroni, N. Bonini, M. Calandra, R. Car, C. Cavazzoni, D. Ceresoli, G. L. Chiarotti, M. Cococcioni, I. Dabo, A. Dal Corso, S. de Gironcoli, S. Fabris, G. Fratesi, R. Gebauer, U. Gerstmann, C. Gougoussis, A. Kokalj, M. Lazzeri, L. Martin-Samos, N. Marzari, F. Mauri, R. Mazzarello, S. Paolini, A. Pasquarello, L. Paulatto, C. Sbraccia, S. Scandolo, G. Sclauzero, A. P. Seitsonen, A. Smogunov, P. Umari, R. M. Wentzcovitch, QUANTUM ESPRESSO: a modular and open-source software project for quantum simulations of materials. *Journal of Physics: Condensed Matter*. **21**, 395502 (2009).
51. M. Schlipf, F. Gygi, Optimization algorithm for the generation of ONCV pseudopotentials. *Computer Physics Communications*. **196**, 36–44 (2015).
52. C. Freysoldt, J. Neugebauer, C. G. van de Walle, Fully *Ab Initio* Finite-Size Corrections for Charged-Defect Supercell Calculations. *Physical Review Letters*. **102**, 016402 (2009).
53. J. P. Perdew, K. Burke, M. Ernzerhof, Generalized Gradient Approximation Made Simple. *Physical Review Letters*. **77**, 3865–3868 (1996).
54. Y. Jin, M. Govoni, G. Wolfowicz, S. E. Sullivan, F. J. Heremans, D. D. Awschalom, G. Galli, Photoluminescence spectra of point defects in semiconductors: Validation of first-principles calculations. *Phys. Rev. Materials*. **5**, 84603 (2021).
55. L. Razinkovas, M. W. Doherty, N. B. Manson, C. G. de Walle, A. Alkauskas, Vibrational and vibronic structure of isolated point defects: The nitrogen-vacancy center in diamond. *Phys. Rev. B*. **104**, 45303 (2021).
56. A. Togo, I. Tanaka, First principles phonon calculations in materials science. *Scripta Materialia*. **108**, 1–5 (2015).

57. H.-Y. Chen, M. Palummo, D. Sangalli, M. Bernardi, Theory and Ab Initio Computation of the Anisotropic Light Emission in Monolayer Transition Metal Dichalcogenides. *Nano Lett.* **18**, 3839–3843 (2018).
58. P. Giannozzi, O. Andreussi, T. Brumme, O. Bunau, M. B. Nardelli, M. Calandra, R. Car, C. Cavazzoni, D. Ceresoli, M. Cococcioni, N. Colonna, I. Carnimeo, A. D. Corso, S. de Gironcoli, P. Delugas, R. A. DiStasio, A. Ferretti, A. Floris, G. Fratesi, G. Fugallo, R. Gebauer, U. Gerstmann, F. Giustino, T. Gorni, J. Jia, M. Kawamura, H.-Y. Ko, A. Kokalj, E. Küçükbenli, M. Lazzeri, M. Marsili, N. Marzari, F. Mauri, N. L. Nguyen, H.-V. Nguyen, A. Otero-de-la-Roza, L. Paulatto, S. Poncé, D. Rocca, R. Sabatini, B. Santra, M. Schlipf, A. P. Seitsonen, A. Smogunov, I. Timrov, T. Thonhauser, P. Umari, N. Vast, X. Wu, S. Baroni, Advanced capabilities for materials modelling with Quantum ESPRESSO. *J. Phys. Condens. Matter.* **29**, 465901 (2017).
59. P. Giannozzi, O. Baseggio, P. Bonfà, D. Brunato, R. Car, I. Carnimeo, C. Cavazzoni, S. de Gironcoli, P. Delugas, F. F. Ruffino, A. Ferretti, N. Marzari, I. Timrov, A. Urru, S. Baroni, Quantum ESPRESSO toward the exascale. *J. Chem. Phys.* **152**, 154105 (2020).
60. L. Patrick, W. J. Choyke, Static Dielectric Constant of SiC. *Phys. Rev. B.* **2**, 15 (1970).
61. Y. Goldberg, M. Levinshtein, S. Rumyantsev, Properties of Advanced Semiconductor Materials: GaN, AlN, InN, BN, SiC, SiGe (John Wiley & Sons, New York, 2001).
62. M. Govoni, G. Galli, Large Scale GW Calculations. *J. Chem. Theory Comput.* **11**, 2680–2696 (2015).

**Acknowledgments**
We thank Grant T. Smith, and Benjamin S. Soloway for helpful discussions. This work made use of the UChicago MRSEC (NSF DMR-1420709) and Pritzker Nanofabrication Facility, which receives support from the SHyNE, a node of the NSF's National Nanotechnology Coordinated Infrastructure (NSF ECCS-1542205). C.P.A., E.O.G., A.B., A.L.C. and D.D.A. were supported by grant nos. AFOSR FA9550-19-1-0358, DARPA D18AC00015KK1932 and ONR N00014-17-1-3026. C.P.A., E.O.G., and D.D.A. were partially supported by the U.S. Department of Energy, Office of Basic Energy Sciences, Materials Science and Engineering Division. C.Z. was supported by Boeing through the Chicago Quantum Exchange. T.O. and H.A. were supported by JPS KAKENHI (grant nos. 21H04553 and 20H00355). J.U.H was supported by the Swedish Research Council (Grant No. 20200544). J.U.H. and N.T.S. were also supported by the EU H2020 FETOPEN project QuanTELCO (grant no. 862721) and the Knut and Alice Wallenberg Foundation (grant no. KAW 2018.0071). Computational work was supported by grant no. AFOSR FA9550-19-1-0358 and made use of computational resources provided by the University of Chicago's Research Computing Center. This work is supported by the U.S. Department of Energy Office of Science National Quantum Information Science Research Centers.

**Author contributions**
C.P.A., E.O.G., and A.B. conceived the experiments. C.P.A., E.O.G., and C.Z. performed the measurements. C.P.A. and E.O.G. analyzed the data. C.P.A. fabricated the devices. A.L.C. assisted in device fabrication. A.B. developed the experimental setup. Y.J. and Y.Z. performed DFT calculations for the energetics and cross-sections. C.V. implemented the calculation of optical matrix elements and assisted with DFT calculations. H.A. and T.O. performed the electron irradiation. J.U.H. and N.T.S. designed and grew the isotopically purified SiC samples. G. G. advised on the computational work. D.D.A. advised on all efforts. All authors contributed to manuscript preparation.

**Data and materials availability**
All data are available in the main text or the supplementary materials. Additional data related to this paper may be requested from the authors.